\documentclass[hyper,letterpaper,12pt]{article}
\usepackage{mathrsfs}

\usepackage{amsfonts}
\usepackage{CJK}
\usepackage{a4wide}
\usepackage{amsmath}
\RequirePackage[dvips]{graphicx}
\usepackage[colorlinks=true,  
            citecolor=red,
            urlcolor= blue,                     %
            filecolor=black,
            linktocpage=true,  
            linkcolor=blue,
             ]{hyperref}

\newcommand{\beq}{\begin{equation}}
\newcommand{\eeq}{\end{equation}}
\newcommand{\bq}{\begin{equation}}
\newcommand{\eq}{\end{equation}}
\newcommand{\ba}{\begin{array}}
\newcommand{\ea}{\end{array}}
\newcommand{\beqa}{\begin{eqnarray}}
\newcommand{\eeqa}{\end{eqnarray}}
\def\bc{\begin{center}}
\def\ec{\end{center}}
\def\bnum{\begin{enumerate}}
\def\enum{\end{enumerate}}
\def\nn{\nonumber}

\renewcommand{\emph}{\normalsize}
\def\Cutoff{Cutoff}
\def\cutoff{cutoff}

\def\hr{\hat{r}}
\def\hrR{\tilde{R}R}
\def\tR{\tilde{R}}
\def\hx{\hat{x}}
\def\hy{\hat{y}}
\def\htau{\hat{\tau}}

\def\hs{{s}}

\def\hu{u}
\def\hK{\hat{K}}
\def\hn{n}
\def\kap{\kappa_4}
\def\etaA{\eta_A}
\def\zetA{\zeta_A}
\def\tetaA{\eta_A}
\def\tzetA{\zeta_A}
\def\teta{\eta}
\def\tbeta{\beta}
\def\tpart{\partial}
\def\tphi{\theta}
\def\tF{F}
\def\tFA{F_A}
\def\tW{\hat{W}}
\def\tN{\hat{N}}
\def\tw{{w_0}}
\def\tp{{p_0}}
\def\te{{e_0}}
\def\ts{{s_0}}
\def\tO{\Omega}
\def\grt{g_{r\hat\tau}(r)}
\def\dhr{\delta \hr}
\def\pa{\partial}
\def\ep{\epsilon}
\def\etwo{(2)}
\def\oo{{\mathcal O}}
\def\tsigma{\sigma}

\def\tT{{\mathcal{T}}}

\def\tP{{\mathcal{P}}}
\def\tE{{\mathcal{E}}}
\def\mF{{\mathcal{F}}}
\def\Cb{C}

\def\zetAp{\hat{\zeta}_A}

\def\Br {H_{r\hat\tau}(r)}
\def\Ar{H_{\hat\tau\hat\tau}(r)}
\def\Cr{H_{\hat{x}\hat{x}}(r)}

\def\Brp {H'_{r\hat\tau}}
\def\Arp{H'_{\hat\tau\hat\tau}}
\def\Crp{H'_{\hat{x}\hat{x}}}

\def\Arpp{H''_{\tau\tau}}
\def\Crpp{H''_{\hat{x}\hat{x}}}

\def\halpha{\hat{\alpha}}
\def\hbeta{\hat{\beta}}
\def\hgamma{\hat{\gamma}}
\def\qq{\qquad}

\def\nn{\nonumber}

\def\[{\left[}
\def\]{\right]}
\def\({\left(}
\def\){\right)}

\def\>{\rightarrow}

\def\Dslash{\not{\hbox{\kern-4pt $D$}}}
\def\pslash{\not{\hbox{\kern-4pt $p$}}}
\def\qslash{\not{\hbox{\kern-4pt $q$}}}
\def\lv{\not{\hbox{\kern-4pt $L$}}}
\def\lsim{\mathrel{\raise.3ex\hbox{$<$\kern-.75em\lower1ex\hbox{$\sim$}}}}
\def\gsim{\mathrel{\raise.3ex\hbox{$>$\kern-.75em\lower1ex\hbox{$\sim$}}}}
\def\ifmath#1{\relax\ifmmode #1\else $#1$\fi}

\evensidemargin=\oddsidemargin

\begin{document}

\begin{titlepage}
\begin{flushright}
\end{flushright}

\begin{center}
 \vspace*{10mm}

{\LARGE\bf Incompressible Navier-Stokes Equations from Einstein Gravity with Chern-Simons Term}\\

\medskip
\bigskip\vspace{0.6cm}
{ {\large\bf Rong-Gen Cai, Tian-Jun Li, Yong-Hui Qi, Yun-Long Zhang
}}
\\[7mm]
{\it  State Key Laboratory of Theoretical Physics, \\
Institute of Theoretical Physics, Chinese Academy of Sciences,\\
Beijing, 100190,  China\\}

\vspace*{0.3cm}
 {\tt cairg@itp.ac.cn,~tli@itp.ac.cn,~qiyh@itp.ac.cn,~zhangyl@itp.ac.cn}
\bigskip\bigskip\bigskip

{
\centerline{\large\bf Abstract}
\begin{quote}
In (2+1)-dimensional hydrodynamic systems with broken parity, the shear and bulk viscosity is joined by the Hall viscosity and
curl viscosity. The dual
holographic model has been constructed by coupling a pseudo scalar
to the gravitational Chern-Simons term in (3+1)-dimensional bulk
gravity. In this paper, we investigate the non-relativistic fluid
with Hall viscosity and curl viscosity living on a finite radial
cutoff surface in the bulk. Employing the non-relativistic
hydrodynamic expansion method, we obtain the incompressible
Navier-Stokes equations with Hall viscosity and curl viscosity.
Unlike the shear viscosity, the ratio of the Hall viscosity over
entropy density is found to be cutoff scale dependent, and it
tends to zero when the cutoff surface approaches to the horizon
of the background spacetime.
\end{quote}}
\end{center}
\end{titlepage}

\tableofcontents

\section{Introduction}

Over the past years there have been a lot of studies on the
gauge/gravity
dualities~\cite{Maldacena:1997re,Gubser:1998bc,Witten:1998qj}. In
the long-wavelength limit, the fluid/gravity correspondence relates
the boundary hydrodynamic equations to the perturbation equations in
the bulk gravity~\cite{Bhattacharyya:2008jc}. Recently, the
(2+1)-dimensional system with broken parity has attracted much
attention, and it has been holographically realized through coupling a pseudo
scalar to the gravitational Chern-Simons term in the (3+1)-dimensional bulk gravity~\cite{Saremi:2011ab,Delsate:2011qp,Chen:2011fs,Jensen:2011xb,Leigh:2012jv,Chen:2012ti}.
The dual gravity model is a generalization of the Chern-Simons modified gravity~\cite{Jackiw:2003pm,Grumiller:2008ie}, which has been widely studied
in the context of cosmology, gravitational waves, and gravitational tests in
the solar and earth systems, see e.g., ~\cite{Smith:2007jm,Grumiller:2007rv}.
In the dual model, $\hrR$ is the parity-violating Pontryagin density~\cite{Deser:1982vy,Deser:1981wh}, which
is a measure of the divergence of axial vector current of massless
chiral fermions $\nabla_{\mu}j^{5\mu }$ in the gravitational anomaly~\cite{AlvarezGaume:1983ig}.

The Hall viscosity $\eta_A$ is a non-dissipative viscosity
coefficient analogous to Hall conductivity which exists in the
(2+1)-dimension fluid with time reversal symmetry breaking. It does not contribute
to the entropy production of the fluid. In the quantum Hall fluids,
the ordinary dissipative viscosities $\eta$ and $\zeta$ are absent
at zero temperature, while the non-dissipative Hall viscosity
 $\eta_A$ could exist~\cite{Avron:1995fg}.
It can be characterized as a rational number and provides a
fundamental measure of incompressibility of the Hall fluid, and it
is related to a Berry phase and proportional to the density of
intrinsic angular
momentum~\cite{{WenZee:1992},Avron:1997,Read:2008rn,Haldane:2009ke,Read:2010,Nicolis:2011ey}.
It is also proposed that the Hall viscosity can be measured as the
coefficient in front of $q^2$ term in the Hall conductivity in small
wave number $q$ limit~\cite{Hoyos:2011ez}.

In the non-relativistic limit of the (2+1)-dimensional hydrodynamics,
the incompressible Navier-Stokes equations will be corrected by the
parity odd viscosities, and it would be interesting to deduce the equations of motion holographically.
In this paper, we use the non-relativistic fluid expansion approach to study the (2+1)-dimensional non-relativistic hydrodynamics with Hall viscosity and curl viscosity,
dual to gravity involving a pseudo scalar coupled to a topological
gravitational Chern-Simons term. 
By using the non-relativistic hydrodynamic expansion method
associated with a flat {\cutoff} surface, we formally solve the
gravity equations and the pseudo scalar equation up to second order
in the non-relativistic hydrodynamic expansion parameter. The
Brown-York stress tensor on the {\cutoff} surface is identified with
the stress energy tensor of the dual fluid, from which we can read
out the general analytic expression for transport coefficient, such
as shear viscosity $\eta$, Hall viscosity $\etaA$ and curl viscosity
$\zetA$. The incompressible Navier-Stokes equations with Hall
viscosity and curl viscosity are obtained from the gravity side. In
addition, it is shown that the Hall viscosity over entropy density
is {\cutoff} scale dependent, and tends to zero when the {\cutoff}
surface approaches the horizon of the background spacetime.

This paper is organized as follows. In section \ref{sec:Fluid}, the
(2+1)-dimensional hydrodynamics with Hall viscosity and bulk
viscosity in non-relativistic limit is investigated. Section
\ref{sec:Fluid-CS} is the main part of this paper, where the
incompressible Navier-Stokes equations with corrections are deduced
from (3+1)-dimensional dual gravity. The case when the cutoff
surface goes to the anti-de Sitter boundary is discussed in section
\ref{sec:boundary}. And some relevant further discussions are given
in section \ref{sec:concl}. In appendix \ref{sec:gs}, the
holographic non-relativistic expansion procedure associated with a
finite {\cutoff} surface in (d+1)-dimensional gravity is briefly
introduced. In this paper, we use the small letter
$i,j,=1,2,...,d-1$ to denote the index of pure spatial coordinate
$x^{i}$, the Greek symbols $\mu,\nu,= 0,1,...,d-1$, to denote index
of the ordinary space-time coordinate $x^{\mu}\sim (\tau,x^{i})$,
and the capital letter $M,N,=r,0,1,...,d-1$ to denote the index of
bulk space-time coordinates $x^{M}\sim (r,x^{\mu})$.

\section{Hydrodynamics with Hall Viscosity and Curl Viscosity}
\label{sec:Fluid}

In a (2+1)-dimensional parity violating hydrodynamic system, the
energy-momentum tensor of the fluid with the first order gradient
expansion can be written
as~\cite{Saremi:2011ab,Chen:2011fs,Jensen:2011xb,Leigh:2012jv,Chen:2012ti}
\beqa T^{\mu\nu}=e u^\mu u^\nu+p
P^{\mu\nu}-2\eta\sigma^{\mu\nu}-\zeta\Theta
P^{\mu\nu}-2\eta_A\sigma^{\mu\nu}_A-\zeta_A\tO P^{\mu\nu}, \eeqa
where the thermodynamic quantities $e$ and $p$ are the energy
density and pressure respectively, which  depend on the local
temperature of the fluid and relate to each other through the
equation of  state $p=p(e)$.
The \emph{shear viscosity} $\teta$ and \emph{bulk viscosity} ${\zeta}$ are the canonical 
transport coefficients, while the \emph{Hall viscosity} $\teta_A$
and \emph{curl viscosity} ${\zeta_A}$ arise from the parity
violating effect. In a (2+1)-dimensional  flat space-time
background, the velocity $u^{\mu}=(1,~\beta^i)/\sqrt{1-\beta^2}$ and
the projection tensor $P^{\mu\nu}=\eta^{\mu\nu}+u^{\mu}u^{\nu}$
should be functions of the space-time coordinates $x^\mu=(\tau,
x^i)$. The first-order gradient expansion tensors can be expressed
as \beqa \sigma^{\mu\nu}& = &
\frac{1}{2}P^{\mu\alpha}P^{\nu\beta}\(\partial_{\alpha}u_{\beta}+\partial_{\beta}u_{\alpha}-\Theta
P_{\alpha\beta}\),\qq
\Theta=\eta^{\mu\nu}\partial_{\mu}u_{\nu},\\
\sigma_{A}^{\mu\nu}& = & \frac{1}{2}\(\epsilon^{\mu\alpha\beta}u_{\alpha}\sigma^{~\nu}_{\beta}
+\epsilon^{\nu\alpha\beta}u_{\alpha}\sigma^{~\mu}_{\beta}\),
\qq\qq\tO=-\epsilon^{\mu\nu\rho}u_\mu\partial_{\nu}u_{\rho}.
\eeqa
where we have used the convention $\ep^{\tau x y}=1$,
and chosen the Landau frame $T^{\mu\nu}u_\nu=-e u^\mu$.

In the thermal equilibrium state, we assume the fluid at rest has
constant energy density $e_0$ and pressure $p_0$. Through
introducing the heat function per unit volume $w_0 = e_0 + p_0$, we
can define the normalized fluctuations of the thermodynamic
parameters as $\tE=(e-e_0)/w_0$ and $\tP=(p-p_0)/w_0$. In the
non-relativistic hydrodynamic limit, the following scalings
appear~\cite{Bhattacharyya:2008kq,Bredberg:2011jq}, \beqa
\beta^\ep_i=\ep \beta_i (\ep x^i, \ep^2 \tau),\qq \tE^\ep=\ep^2 \tE
(\ep x^i, \ep^2 \tau),\qq \tP^\ep=\ep^2 \tP (\ep x^i, \ep^2
\tau),\label{Eq:NRlimit} \eeqa where $\ep\ll 1$ is a small
parameter. The velocity $\beta_i$ could be regarded as a small
vector fluctuation, while the thermodynamic quantities $\tE$ and
$\tP$ basically come from the temperature fluctuation, and they
would relate to each other through the equation of state $p=p(e)$ of
the fluid. Up to the second order of the non-relativistic expansion
parameter $\ep$, one has \beqa
u_{\mu}&\sim & (-1,~0)+(~0,~\beta_{i})\,\epsilon+\frac{1}{2}({\beta^{2}},~0)\,\epsilon^{2}+\oo(\epsilon^{3}),\\[0.1cm]
P_{\mu\nu}&\sim &\left(
                       \begin{array}{cc}
                        0&0\\
                        0& \delta_{ij}\\
                        \end{array}
                    \right)
             - \left(
                       \begin{array}{cc}
                        0&\beta_{j}\\
                        \beta_{i}& 0\\
                        \end{array}
                    \right)\epsilon
            +  \left(
                       \begin{array}{cc}
                        \beta^{2}&  0\\
                        0 & \beta_{i}\beta_{j}\\
                        \end{array}
                    \right)\epsilon^{2}
            + \oo(\epsilon^{3}).
\eeqa
The non-zero components of the normal traceless shear viscosity tensor
and the traceless Hall viscosity tensor are
\beqa
\sigma_{ij}&\sim & \frac{1}{2}\left(
        \begin{array}{ccc}
        (\partial_{x}\beta_{x}-\partial_{y}\beta_{y})& (\partial_{x}\beta_{y}+\partial_{y}\beta_{x})\\
        (\partial_{x}\beta_{y}+\partial_{y}\beta_{x})& (\partial_{y}\beta_{y}-\partial_{x}\beta_{x})\\
        \end{array}
         \right)   \epsilon^{2}+\oo(\epsilon^{3}) ,\\[0.4cm]
\sigma^{A}_{ij}&\sim & \frac{1}{2}\left(
        \begin{array}{ccc}
        (+\partial_{x}\beta_{y}+\partial_{y}\beta_{x})& (-\partial_{x}\beta_{x}+\partial_{y}\beta_{y})\\
        (-\partial_{x}\beta_{x}+\partial_{y}\beta_{y})& (-\partial_{x}\beta_{y}-\partial_{y}\beta_{x})\\
        \end{array}
         \right)   \epsilon^{2}+\oo(\epsilon^{3}),
\eeqa
the divergence and the curl of the velocity become
\beqa
\Theta= (\partial_{x}\beta_{x}+\partial_{y}\beta_{y})\epsilon^{2}+\oo(\epsilon^{3}),\qq
\tO= (\partial_{x}\beta_{y}-\partial_{y}\beta_{x})\epsilon^{2}+\oo(\epsilon^{3}).
\eeqa
The dynamical equations of the relativistic fluid are
$\partial_\mu T^{\mu\nu}=\mF^{\nu}$, where $\mF^{\nu}$ is an external force density,
and we can define the normalized force density as $f^{\nu}=\mF^{\nu}/w_0$.
In the non-relativistic limit (\ref{Eq:NRlimit}),
$\partial_\mu T^{\mu\nu}=\mF^{\nu}$ will reduce to the incompressible Navier-Stokes equations with Hall viscosity and curl viscosity,
\beqa\label{Eq:NS}
\partial^{i}\beta_{i}=0,\quad \partial_{\tau}\beta_{i}+\beta_{j}\partial^{j}\beta_{i}+\partial_{i}\tP
-{\nu}\,\partial^{2}\beta_{i}-{\nu}_{A}\epsilon^{ij}\partial^{2}\beta_{j}-\xi_{A} \epsilon^{jk}\partial_{i}\partial_{j}\beta_{k}=f_i,
\eeqa
where
${\nu}\equiv\teta/w_0$,
${\nu_A}\equiv\teta_A/w_0$ and
${\xi_A}\equiv\zeta_A/w_0$
could be named as the \emph{kinematic shear viscosity},
\emph{kinematic Hall viscosity} and \emph{kinematic curl viscosity},  respectively.

\section{Parity Breaking Hydrodynamics from Gravity}
\label{sec:Fluid-CS} In this section, we will deduce the
(2+1)-dimensional  non-relativistic parity breaking hydrodynamics described in the
previous section from the (3+1)-dimensional gravity.

\subsection{The Dual Gravity Model}
\label{sec:CSGLagrangian}

The dual gravity model is described by the Einstein gravity
involving a pseudo scalar coupled to the topological gravitational
Chern-Simons term. The action of the model can be expressed
as\footnote{For the surface term, see, e.g.,
~\cite{Grumiller:2008ie,Bakas:2008gz,Miskovic:2009bm}.}~\cite{Saremi:2011ab}
\beqa S_{bulk}=\frac{1}{2\kap}\int d^{4}x\sqrt{-g} (R-2\Lambda+
{\mathcal L}_{CS} + {\mathcal L}_{\theta}),
\label{Eq:CS-gravity}
 \eeqa
 where $g=\det{g_{MN}}$ is the
determinant of the metric, $R$ is the Ricci scalar, $\Lambda$ is the
cosmological constant\footnote{If appropriate solutions exist in
this system, we do not require the cosmological constant to be
negative.}, $\kap=8\pi G_{N}$ with $G_{N}$ the Newton's
gravitational constant. The Lagrangian density ${\mathcal L}_{CS}$
and ${\mathcal L}_{\theta}$ are
 \beqa {\mathcal L}_{CS}=\frac{\lambda }{4}\tphi
\hrR, \qquad {\mathcal
L}_{\theta}=-\frac{1}{2}g^{MN}\nabla_{M}\tphi\nabla_{N}\tphi-V(\tphi),
\label{Eq:LCS-theta} \eeqa
where $\lambda$ is a coupling constant
and $\theta$ is the pseudo scalar field. The Pontryagin density
$\hrR$ is defined as \beqa \hrR=\tR^{M
\,\,PQ}_{\,\,\,\,N}{R}^{N}_{\,\,MPQ}, \quad~~\tR^{M
\,\,PQ}_{\,\,\,\,N} = \frac{1}{2}\epsilon^{PQAB}R^{M}_{\,\,\,\,NAB},
\eeqa where $\epsilon^{MNAB}$ is the four-dimensional Levi-Civita
tensor in the bulk with the convention $\epsilon^{r\tau
xy}=1/\sqrt{-g}$. Varying the action with respect to the metric and
the pseudo scalar respectively leads to the equations of motion
 \beqa
W_{MN}&=& 0,\qq W_{MN}\equiv R_{MN}-\frac{1}{2}g_{MN}R+\Lambda
g_{MN}+\lambda C_{MN}-\kap T_{MN}^{\tphi}, \label{Eq:EMNt} \\
W_{\tphi}&=& 0,\qq~~~ W_{\tphi}\equiv2\kap\(\nabla^{2}\tphi
 -\frac{dV(\tphi)}{d\tphi}\)-\frac{\lambda}{4}\hrR,\label{Eq:theta}
\eeqa
where the stress energy tensor of the pseudo scalar is
\beqa
T_{MN}^{\tphi} &\equiv &
-2\frac{\delta {\mathcal L}_{\theta} }{~\delta g^{MN}}+g_{MN}{\mathcal L }_{\theta}
=\nabla_{M}\tphi\nabla_{N}\tphi-\(\frac{1}{2}g^{AB}\nabla_{A}\tphi\nabla_{B}\tphi+V(\tphi)\)g_{MN}, \label{Eq:TMN-theta}
\eeqa
and the four-dimensional Cotton tensor $C_{MN}$ is a symmetric traceless tensor of the second rank,
which is defined through\footnote{It was also named as $C$-Tensor in~\cite{Saremi:2011ab}.}~\cite{Jackiw:2003pm,Grumiller:2008ie}
\beqa
C^{MN}&=& -v_{S}\epsilon^{SAB(M}\nabla_{A}R^{N)}_{B}
+v_{ST}\tR^{T(MN)S}, \label{Eq:Cmn}\\
v_{S}&=&\nabla_{S}\tphi, \quad v_{ST}=\nabla_{S}\nabla_{T}\tphi =\nabla_{(S}\nabla_{T)}\tphi. \label{Eq:vs}
\eeqa
The Cotton tensor's divergence has a non-zero topological source as
\beqa
\nabla_{M}C^{MN}=\frac{v^{N}}{8}\hrR=\frac{v^{N}}{4}\partial_{M}J^{M},\label{Eq:pCmn}
\eeqa
where $J^M$ origins from the 4-divergence of the gravitational Chern-Simons topological current
\beqa
J^{M}&=& \epsilon^{MNPQ}(\Gamma^{A}_{NB}
\partial_{P}\Gamma^{B}_{QA}+
\frac{2}{3}\Gamma^{A}_{NB}\Gamma^{B}_{PS}\Gamma^{S}_{QA}).
\eeqa

The gravity equations $W_{MN}=0$ simultaneously imply
$\nabla^{M}W_{MN}=0$. Meanwhile, the Bianchi identity
$\nabla_{(L}R_{PQ)MN}=0$ leads to
$\nabla^{M}(R_{MN}-\frac{1}{2}g_{MN}R+\Lambda g_{MN})=0$. Thus,
assuming that the parameters $\lambda$ and $\kap$ are constants, we
have \beqa \kap\nabla^{M}T_{MN}^{\tphi}=\lambda \nabla^{M}C_{MN}.
\label{Eq:Bianchi} \eeqa
On the other hand from Eq.~(\ref{Eq:TMN-theta}) and
Eq.~(\ref{Eq:pCmn}) one has \beqa \nabla^{M}T_{MN}^{\tphi}=v_{N}
\(\nabla^{2}\tphi-\frac{dV(\tphi)}{d\tphi}\),\qq
\nabla^{M}C_{MN}=\frac{v_{N}}{8}\hrR, \label{Eq:Bianchi2} \eeqa from
which together with Eq.~(\ref{Eq:Bianchi}), we arrive at either
$v_{N}=\nabla_N \theta=0$ or $W_{\tphi}=0$.  As we are interested in
the case where the pseudo scalar field $\tphi$ depends on the
space-time coordinates, i.e. $v_{N}\neq 0$, thus, the Bianchi
identity together with Einstein equations leads to the pseudo scalar
equation $W_{\tphi}=0$.

\subsection{Non-relativistic Hydrodynamics Expansion }

From the above discussion we can see that the pseudo scalar equation
is not independent. Thus, we will focus on the gravity equations
henceforth. The Ricci scalar can be obtained from the trace of
Eq.~(\ref{Eq:EMNt}) as \beqa R-4\Lambda+\lambda
C^M_{\,M}-\kap[(\partial\tphi)^{2}+4V(\tphi)]=0, \label{Eq:RicciS}
\eeqa which leads to the trace-reversed form of the gravity
equations: $E_{MN}+\lambda c_{MN}=0$, where \beqa E_{MN}\equiv
R_{MN}-\Lambda g_{MN}-\kap t_{MN},\label{Eq:EMNt2} \qq t_{MN} \equiv
\partial_{M}\tphi\partial_{N}\tphi+g_{MN}V(\tphi). \eeqa The
simplified Cotton tensor $c_{MN}=C_{MN}$ is due to  its traceless
property $C^{M}_{\,\,M}=0$. Then the equations of motion (EOMs) to
be solved are \footnote{In what follows, we will use the
normalization $2 \kap\equiv 1$, which means $16\pi G_{N} \equiv 1$.}
\beqa
\tW_{MN}&=&0,\qq\tW_{MN}\equiv E_{MN}+\lambda C_{MN}, \label{Eq:WMN}\\
\qq \tW_{\tphi}&=&0,\qq~~~\tW_{\tphi}\equiv \nabla^{2}\tphi
-\frac{dV(\tphi)}{d\tphi}-\frac{\lambda}{4}\hrR. \label{Eq:Wth}
\eeqa We assume that the following general (3+1)-dimensional black
brane metric \beqa
ds^{2}&=&-g_{tt}(r)dt^{2}+g_{rr}(r)dr^{2}+g_{xx}(r)\delta_{ij}dx^{i}dx^{j},\quad
i,j=1,2,\label{Eq:ds2-generic} \eeqa as well as a radial coordinate
dependent pseudo scalar field $\tphi(r)$ solves the EOMs. With
Dirichlet boundary condition at the finite {\cutoff} surface
$r=r_c$, keeping the induced metric flat, and employing the
non-relativistic hydrodynamics expansion method  (see appendix
\ref{sec:gs} for details), one can perturb the background metric up
to the second order of expansion parameter $\epsilon\ll 1$, \beqa
&d\hs^{2}_{(b)}&= g_{MN}dx^{M}dx^{N}=  2\grt drd\htau -\frac{g_{tt}(r)}{g_{tt}(r_{c})}d\htau^{2}+\frac{g_{xx}(r)}{g_{xx}(r_c)}\delta_{ij}d\hx^{i}d\hx^{j}\nn\\
&&~-2\grt \tbeta_{i}d\hx^{i}dr+2\(\frac{g_{tt}(r)}{g_{tt}(r_c)}-\frac{g_{xx}(r)}{g_{xx}(r_c)}\)\tbeta_{i}d\hx^{i}d\htau\nn\\
&&~+\grt \tbeta^{2}drd\htau-\(\frac{g_{tt}(r)}{g_{tt}(r_c)}-\frac{g_{xx}(r)}{g_{xx}(r_c)}\)\(\tbeta^{2}d\htau^{2}+\tbeta_{i}\tbeta_{j}d\hx^{i}d\hx^{j}\)\nn\\
&&~+ \(\dhr\)\grt \frac{g'_{rr}(r)}{g_{rr}(r)}drd\htau
+\(\dhr\)\(\frac{g'_{tt}(r)}{g_{tt}(r)}-\frac{g'_{tt}(r_c)}{g_{tt}(r_c)}\)\(\grt drd\htau -\frac{g_{tt}(r)}{g_{tt}(r_{c})}d\htau^{2}\)\nn\\
&&~+\(\dhr\)\(\frac{g'_{xx}(r)}{g_{xx}(r)}-\frac{g'_{xx}(r_c)}{g_{xx}(r_c)}\)\frac{g_{xx}(r)}{g_{xx}(r_c)}\delta_{ij}d\hx^{i}d\hx^{j}+\oo(\epsilon^3),\label{Eq:ds2-nc}
\eeqa where we have defined $\grt\equiv\sqrt{g_{rr}(r)}\sqrt{
{g_{tt}(r)} / {g_{tt}(r_c)}}$. The bulk coordinates are
$x^M=(r,\hx^\mu)$, the intrinsic coordinates
$\hx^\mu=(\htau,\hx^i)=(\htau,\hx,\hy)$ and we denote
$\tpart_{i}\equiv \pa_{\hx^i}$\footnote{Here and henceforth, the
subscript $i$ denotes $\hx$ or $\hy$.}. The pseudo scalar field
should also be expanded up to the second order \beqa
\hat\tphi(r)\equiv\tphi(\hr)=\tphi(r)+\tphi'(r)\dhr. \eeqa We have
assumed that the perturbation parameters are intrinsic coordinates
$\hx^\mu$ dependent, where $\dhr=\hr-r=\dhr(\hx^\mu)$ is a scalar
perturbation at order $\ep^2$
and $\beta^{i}=\beta^{i}(\hx^\mu)$ is 
the Lorentz boost parameter at order $\ep^1$. Together with the
non-relativistic low frequency, long wavelength limit, we use the
following scalings
\beqa
\partial_{r}\sim \epsilon^{0}, \quad \tpart_{i}\sim\tbeta_{i}(\hx^\mu)\sim \epsilon^{1},\quad
\tpart_{\htau}\sim \delta \hr(\hx^\mu)\sim\epsilon^{2}. \label{Eq:HDE}
\eeqa
Then  we can solve the EOMs order by order with the non-relativistic hydrodynamics expansion.

The Cotton tensor firstly presents at order $\epsilon^{2}$, namely
$C_{MN}^{(0)}\equiv 0$ and $C_{MN}^{(1)}\equiv 0$. Thus we obtain background equations
\beqa
E_{rr}^{(0)}=0,\quad E_{\htau\htau}^{(0)}=0,\quad  E_{ii}^{(0)}=0,\quad \tW^{(0)}_{\tphi}=0, \label{Eq:Bg}
\eeqa
where three of them are independent, and
\begin{align}
E_{rr}^{(0)}&=\frac{1}{2}\(\frac{g_{rr}^{\prime}(r)}{g_{rr}(r)}+\frac{g_{tt}^{\prime}(r)}{g_{tt}(r)}+\frac{g_{xx}^{\prime}(r)}{g_{xx}(r)}\)\frac{g_{xx}^{\prime}(r)}{g_{xx}(r)}-\frac{1}{2}\tphi^{\prime 2}(r) -\frac{g_{xx}^{\prime\prime}(r)}{g_{xx}(r)},\label{Err}\\
E_{\htau\htau}^{(0)}&= \frac{g_{tt}(r)}{g_{tt}(r_c)}\(\Lambda +\frac{1}{2}V(\tphi(r))+\frac{1}{2g_{rr}(r)}\frac{g_{tt}^{\prime}(r)}{g_{tt}(r)}\[\frac{g_{tt}^{\prime\prime}(r)}{g_{tt}^{\prime}(r)}+\frac{g_{xx}^{\prime}(r)}{g_{xx}(r)}-\frac{1}{2}\(\frac{g_{rr}^{\prime}(r)}{g_{rr}(r)}+\frac{g_{tt}^{\prime}(r)}{g_{tt}(r)}\)\] \),\label{Ett}\\
E_{ii}^{(0)}&= -\frac{g_{xx}(r)}{g_{xx}(r_c)}\(\Lambda +\frac{1}{2}V(\tphi(r))+\frac{1}{2g_{rr}(r)}\frac{g_{xx}^{\prime}(r)}{g_{xx}(r)}\[\frac{g_{xx}^{\prime\prime}(r)}{g_{xx}^{\prime}(r)}-\frac{1}{2}\(\frac{g_{rr}^{\prime}(r)}{g_{rr}(r)}-\frac{g_{tt}^{\prime}(r)}{g_{tt}(r)}\)\]\),\label{Eii} \\
\tW^{(0)}_{\tphi}&=\frac{\tphi^{\prime\prime}(r)}{g_{rr}(r)}
+\(\frac{g_{xx}^{\prime}(r)}{g_{xx}(r)}+\frac{g_{tt}^{\prime}(r)}{2g_{tt}(r)}
-\frac{g_{rr}^{\prime}(r)}{2g_{rr}(r)}\)\frac{\tphi^{\prime}(r)}{g_{rr}(r)}
-\frac{dV(\tphi(r))}{d\tphi(r)}.\label{Escalar}
\end{align}
%
We assume that background equations are always satisfied and they can be used to simplify the perturbation equations at higher order.
It turns out that the intrinsic coordinate dependent metric and
pseudo scalar can only solve the EOMs (\ref{Eq:WMN}) and
(\ref{Eq:Wth}) up to order $\epsilon^{1}$. Employing the spatial
rotational $SO(2)$ symmetry between $\hx^i$ directions, the nonzero traceless
tensor sector and scalar sector appear in $\tW_{MN}$ firstly at order $\ep^2$.
Correction terms to the metric and pseudo scalar are required to
cancel the corresponding  structure. For the traceless shear tensor
sector, the following tensor correction terms are needed  in the
metric at order $\epsilon^2$
\beqa
d\hs^{2}_{(t,\ep^2)}&=&\frac{g_{xx}(r)\sqrt{g_{tt}(r_{c})}}{g_{xx}(r_c)}
\(2\tF(r)~\tsigma_{ij}+2\tFA(r)~\tsigma^{A}_{ij}\)d\hx^{i}d\hx^{j},\\
\tsigma_{ij}&=&\tpart_{(i}\tbeta_{j)}-\frac{1}{2}\delta_{ij}\tpart_{k}\tbeta^{k},\qq
\label{Eq:ds2Fr}
\tsigma^{A}_{ij}=\frac{1}{2}\(\epsilon_{ik}\tsigma^{~k}_{j}+\epsilon_{jk}\tsigma^{~k}_{i}\),
\label{Eq:ds2Gr} \eeqa where $F(r)$ and $\tFA(r)$ are chosen to
cancel the ordinary \emph{shear tensor} $\tsigma_{ij}$ and
\emph{Hall tensor} $\tsigma^{A}_{ij}$ of source terms at order
$\epsilon^{2}$,  respectively.

 For the scalar sector,
there are four perturbation equations. The Bianchi identity leads to
a constraint among them, thus, only three are independent. Although
one can add three  corrections terms in the metric and one in the
pseudo scalar, a gauge choice for these perturbations is needed. We
choose a gauge  so that no corrections appear in the pseudo scalar,
and introduce the following three correction terms in the metric to
cancel the residual \emph{curl scalar} $\tO$ appearing in the Cotton
tensor at order $\epsilon^{2}$
 \beqa
d\hs^{2}_{(s,\ep^2)}&=&\(2\grt \Br drd\htau -\frac{g_{tt}(r)}{g_{tt}(r_c)}\Ar d\htau^{2}+\frac{g_{xx}(r)}{g_{xx}(r_c)}\Cr d\hx_{i}d\hx^{i}\)\tO,  \label{Eq:ds2ABCr}\\
\tO & \equiv & \epsilon^{ij}\tpart_{i}\tbeta_{j}=\tpart_{\hx}\tbeta_{\hy}-\tpart_{\hy}\tbeta_{\hx}.
\label{Eq:Hallbulk}
\eeqa

In summary, to solve the trace-reversed form of the gravity equations
$\tW_{MN}=0$ and the pseudo scalar equation ${\tW}_{\tphi}=0$,
we will work with the perturbed metric and pseudo scalar field with corrections
up to $\ep^2$ in the non-relativistic hydrodynamic limit,
\beqa
d\hs^{2}&=&d\hs^{2}_{(b)}+d\hs^{2}_{(t,\ep^2)}+d\hs^{2}_{(s,\ep^2)},\label{Eq:fullmetric}\\
\hat\tphi(r)&=&\tphi(r)+\tphi'(r)\dhr, \label{Eq:fullscalar} \eeqa
where the subscripts $(b),(t),(s)$ in the metric represent
``background", ``tensor", and ``scalar",  respectively.
 $\hat\tphi(r)$
is substituted for $\tphi$ in EOMs. In addition, let us mention here
that correction terms to the vector sector appear at order $\ep^3$,
thus will not be considered in this paper.

\subsubsection{Traceless Tensor Sector of the Perturbation Equations}

By substituting the metric (\ref{Eq:fullmetric}) and pseudo scalar
field (\ref{Eq:fullscalar}) into the EOMs $\tW_{MN}=0$, the
traceless tensor sector gives the following second order ordinary
differential equation \beqa
0&=&\frac{d}{dr}\[g_{xx}(r)\(\frac{\sqrt{g_{tt}(r)}}{\sqrt{g_{rr}(r)}}F'(r)+1\)\]\sigma_{ij}\nn\\
&+&\frac{d}{dr}\[g_{xx}(r)\(\frac{\sqrt{g_{tt}(r)}}{\sqrt{g_{rr}(r)}}F_A'(r)+\frac{\lambda}{2g_{rr}(r)}\(\frac{g_{tt}^{\prime}(r)}{g_{tt}(r)}-\frac{g_{xx}^{\prime}(r)}{g_{xx}(r)}\)\tphi^{\prime}(r)\)\]\sigma^{A}_{ij}.
\eeqa As $\sigma_{ij}$ and $\sigma^{A}_{ij}$ have different tensor
structure, we can solve the two second order differential equations,
respectively. Here we are interested in the physics between the black
brane horizon $\Sigma_h$ and the {\cutoff} surface $\Sigma_c$. At
the {\cutoff} surface $\Sigma_c$, we impose the Dirichlet boundary
condition to keep the induced metric $\hgamma_{\mu\nu}(r_c)$ flat,
i.e., $F(r_{c})=0$ and  $F_A(r_{c})=0$. At the horizon $r=r_h$, we
demand  $\tF(r_{h})$ and $\tFA(r_{h})$ to be finite. Then the
function $F(r)$ and  $F_A(r)$ can be determined as
\beqa \tF(r)&=&
\int_{r}^{r_c}dy\sqrt\frac{g_{rr}(y)}{g_{tt}(y)}\[1-\frac{c_{F}}{g_{xx}(y)}\],\\
\label{Eq:Fr} \tFA(r)&=&
\int_{r}^{r_c}dy\sqrt\frac{g_{rr}(y)}{g_{tt}(y)}\[\frac{\lambda}{2g_{rr}(y)}\(\frac{g_{tt}^{\prime}(y)}{g_{tt}(y)}-\frac{g_{xx}^{\prime}(y)}{g_{xx}(y)}\)\tphi^{\prime}(y)-\frac{c_{F_A}}{g_{xx}(y)}\],
\label{Eq:FrA} \eeqa where the two integration constants are
 \beqa
c_F&=& {g_{xx}(r_h)},\\ \label{Eq:cFr}
c_{F_A}&=&\frac{\lambda}{2}\frac{g_{xx}(r_h)}{g_{rr}(r_h)}\(\frac{g_{tt}^{\prime}(r_h)}{g_{tt}(r_h)}-\frac{g_{xx}^{\prime}(r_h)}{g_{xx}(r_h)}\)\tphi^{\prime}(r_h)
\label{Eq:cFrA}
=\frac{\lambda}{2}\frac{g_{xx}(r_h)}{g_{rr}(r_h)}\frac{g_{tt}^{\prime}(r_h)}{g_{tt}(r_h)}\tphi^{\prime}(r_h).
\label{Eq:cFrA} \eeqa Here we have used the assumption that at the
horizon of the black brane solution (\ref{Eq:ds2-generic}),
$g_{tt}(r)$ has the first order zero $g_{tt}(r_h)=0$, and
$g_{rr}(r)$ has the first order pole $g^{-1}_{rr}(r_h)=0$,
while their product $g_{tt}(r_h)g_{rr}(r_h)$ is finite.

\subsubsection{Scalar Sector of the Perturbation Equations}
After solving the traceless tensor perturbation equations, only
scalar sectors are left at order $\ep^2$. At first, it is worthy to
emphasize that the \emph{constraint equations} on the \emph{scalar
sector} of gravity equations at order $\ep^2$ give \beqa \tN^M
\tW_{M\htau}=\frac{1}{2}\tw(r)\tpart_i \beta^i=0,\qq
\tw(r)=\frac{1}{\sqrt{g_{rr}(r)}}\(\frac{g'_{tt}(r)}{g_{tt}(r)}-\frac{g'_{xx}(r)}{g_{xx}(r)}\),
\label{Eq:cons-2} \eeqa where $\tN$ is the unit normal vector of
the constant $r$ surface. As $\tw(r)$ is not identically equal to
zero, the constraint equation (\ref{Eq:cons-2}) leads to the
incompressibility condition $\Theta=\tpart_i \beta^i=0$, which shows
the velocity field is divergence-free while the vorticity
$\tO=\ep^{ij}\tpart_i \beta_j$ is still allowed.

Then we need to solve scalar sector of the perturbation equations in (\ref{Eq:WMN})
 \beqa &&\tW^{\etwo}_{rr}=0,\quad
\tW^{\etwo}_{\htau\htau}=0,\quad
\tW^{\etwo}_{\hx\hx}+\tW^{\etwo}_{\hy\hy}=0, \quad \tW^{(2)}_{\tphi}=0.\label{Eq:Scalarp} \eeqa
Among them, only three are independent.
Using the solutions (\ref{Eq:fullmetric}) and (\ref{Eq:fullscalar}),
we can obtain the scalar sector of gravity equations
defined in (\ref{Eq:EMNt2}) at order  $\epsilon^{2}$,
%
\begin{align}
&{E_{rr}^{\etwo}}= \[\frac{g_{xx}^{\prime}(r)}{g_{xx}(r)}\Brp(r)+\(\frac{g_{rr}^{\prime}(r)}{2g_{rr}(r)}+\frac{g_{tt}^{\prime}(r)}{2g_{tt}(r)}-\frac{g_{xx}^{\prime}(r)}{g_{xx}(r)}\)\Crp(r)-\Crpp(r)\]{\tO},\nn\\
&{E_{\hat\tau\hat\tau}^{\etwo}}= \[\frac{g_{tt}^{\prime}(r)}{g_{tt}(r_c)}\frac{\Crp(r)-\Brp(r)}{2g_{rr}(r)}+\(\frac{g_{tt}(r)}{g_{tt}(r_c)}\(\frac{g_{xx}^{\prime}(r)}{g_{xx}(r)}-\frac{1}{2}\frac{g_{rr}^{\prime}(r)}{g_{rr}(r)}\)+\frac{3}{2}\frac{g_{tt}^{\prime}(r)}{g_{tt}(r_c)}\)\frac{\Arp(r)}{2g_{rr}(r)}
+\frac{g_{tt}(r)}{g_{tt}(r_c)}\frac{\Arpp(r)}{2g_{rr}(r)}\right.\nn\\
&\qq\left.+\(\frac{g_{tt}(r)}{g_{tt}(r_c)}\frac{g_{xx}^{\prime\prime}(r)}{g_{xx}(r)}+\frac{1}{2}\frac{g_{xx}^{\prime}(r)}{g_{xx}(r)}\(\frac{g_{tt}^{\prime}(r)}{g_{tt}(r_c)}-\frac{g_{tt}(r)}{g_{tt}(r_c)}\frac{g_{rr}^{\prime}(r)}{g_{rr}(r)}\)\)\frac{\Ar -2\Br}{2g_{rr}(r)}\right]{\tO},\label{Eq:E2tt}\nn\\
&{E_{\hx\hx}^{\etwo}+E_{\hy\hy}^{\etwo}}=\[\frac{g_{xx}^{\prime}(r)}{g_{xx}(r_c)}\frac{\Brp(r)-\Arp(r)}{g_{rr}(r)}
+\(\frac{1}{4}\frac{g_{xx}(r)}{g_{xx}(r_c)}\(\frac{g_{rr}^{\prime}(r)}{g_{rr}(r)}-\frac{g_{tt}^{\prime}(r)}{g_{tt}(r)}\)-\frac{g_{xx}^{\prime}(r)}{g_{xx}(r_c)}\)\frac{2\Crp(r)}{g_{rr}(r)}\right.\nn\\
&\qq\left.-\frac{g_{xx}(r)}{g_{xx}(r_c)}\frac{\Crpp(r)}{g_{rr}(r)}+\(\frac{g_{xx}^{\prime}(r)}{2g_{xx}(r_c)}\(\frac{g_{rr}^{\prime}(r)}{g_{rr}(r)}-\frac{g_{tt}^{\prime}(r)}{g_{tt}(r)}\)-\frac{g_{xx}^{\prime\prime}(r)}{g_{xx}(r_c)}\)\frac{\Ar -2\Br}{g_{rr}(r)}\]{\tO},
\end{align}
and we have imposed the incompressibility condition here and
henceforth. The corresponding scalar sector of Cotton tensor defined
in Eq.~(\ref{Eq:Cmn}), appears firstly at order $\epsilon^{2}$, can
be expressed as
\begin{align}
&{C_{rr}^{\etwo}}=\frac{-1}{2\sqrt{g_{rr}(r)}}\sqrt\frac{g_{tt}(r)}{g_{tt}(r_c)}\frac{g_{xx}(r_c)}{g_{xx}(r)}\frac{g_{rr}(r)}{g_{xx}(r)}~\frac{d}{dr}\[\frac{g_{xx}(r)}{g_{rr}(r)}\(\frac{g_{tt}^{\prime}(r)}{g_{tt}(r)}-\frac{g_{xx}^{\prime}(r)}{g_{xx}(r)}\)\tphi^{\prime}(r)\]{\tO},\nn\\
&{C_{\htau\htau}^{\etwo}}=\frac{-1}{2\sqrt{g_{rr}(r)}}\frac{g_{tt}(r)}{g_{tt}^{{3}/{2}}(r_c)}\frac{g_{xx}(r_c)}{g_{xx}^{{3}/{2}}(r)}
~\frac{d}{dr}\[\frac{\sqrt{g_{tt}(r)}\sqrt{g_{xx}(r)}}{g_{rr}(r)}\(\frac{g_{tt}^{\prime}(r)}{g_{tt}(r)}-\frac{g_{xx}^{\prime}(r)}{g_{xx}(r)}\)\tphi^{\prime}(r)  \]{\tO},\nn\\
&{C_{\hx\hx}^{\etwo}+C_{\hy\hy}^{\etwo}}=\frac{-1}{2\sqrt{g_{rr}(r)}}\frac{1}{\sqrt{g_{tt}(r_c)}}\frac{1}{\sqrt{g_{tt}(r)}}~\frac{d}{dr}\[\frac{g_{tt}(r)}{g_{rr}(r)}\(\frac{g_{tt}^{\prime}(r)}{g_{tt}(r)}-\frac{g_{xx}^{\prime}(r)}{g_{xx}(r)}\)\tphi^{\prime}(r)\]{\tO}.
\end{align}
In addition, we have the pseudo scalar equation defined in
Eq.~(\ref{Eq:Wth}) at order $\ep^2$, \beqa
{\tW^{(2)}_{\tphi}}&=&\[\frac{\tphi^{\prime}(r)}{g_{rr}(r)}\(\Arp(r)-\Brp(r)+\Crp(r)\)+\frac{dV(\tphi(r))}{d\tphi(r)}\(\Ar -2\Br \)\right.\nn\\
&&\left.-
\frac{\lambda}{4}\frac{2}{g_{rr}^{{3}/{2}}(r)}\frac{g_{xx}(r_c)}{g_{xx}(r)}\frac{g_{xx}^{\prime}(r)}{g_{xx}(r)}\sqrt\frac{g_{tt}(r)}{g_{tt}(r_c)}\(\frac{g_{tt}^{\prime}(r)}{g_{tt}(r)}-\frac{g_{xx}^{\prime}(r)}{g_{xx}(r)}\)^{2}\]\tO.
\label{Eq:Scalar_2} \eeqa Now we have three functions to be
determined through four equations, for the latter, only three of
them are independent due to the Bianchi identity. With the Dirichlet
boundary condition at the cutoff surface, namely
$H_{\htau\htau}(r_c)=0$, $H_{\hx\hx}(r_c)=0$ and the boundary
condition at the horizon,  in principle we can determine the three
functions $\Ar $, $\Br $ and $\Cr $ in terms of the background
metric functions $g_{tt}(r)$, $g_{rr}(r)$ and $g_{xx}(r)$. But due
to the complexity of equations, here we do not intend to present
these expressions explicitly. Instead we just assume that the three
functions $\Ar $, $\Br $ and $\Cr $ solve the three independent
scalar perturbation equations in (\ref{Eq:Scalarp}), as  the
assumption that the black brane metric (\ref{Eq:ds2-generic}) solves
the background equations of motion~\cite{Gubser:2008sz}.

\subsection
{Incompressible Navier-Stokes at the {\Cutoff} Surface}
 \label{sec:}

We now study the (2+1)-dimensional hydrodynamics in the
non-relativistic limit dual to the background gravity solution. At
the {\cutoff} surface $\Sigma_c$, the Brown-York
tensor~\cite{Brown:1992br} \beqa
\tT_{\mu\nu}=2\(\hK\hgamma_{\mu\nu}(r_c)-\hK_{\mu\nu}+\Cb\hgamma_{\mu\nu}(r_c)\)
\label{Eq:Tmn} \eeqa is identified with the stress energy of the
dual fluid in Ref.~\cite{Bredberg:2010ky}, where $\hK_{\mu\nu}$ is
the extrinsic curvature tensor,
$\hgamma_{\mu\nu}(r_c)=\eta_{\mu\nu}$ is the induced metric on the
{\cutoff} surface, and $C$ is a constant. In the non-relativistic
limit, the non-zero components at order $\epsilon^{0}$ turn out to
be \beqa
\tT_{\htau\htau}^{ (0)}&=&\te(r_c),\qq~~~\te(r_c)=-\frac{2}{\sqrt{g_{rr}(r_c)}}\frac{g_{xx}^{\prime}(r_c)}{g_{xx}(r_c)}-2\Cb,\\
 \tT_{ij}^{(0)}&=&\tp(r_c)\delta_{ij},\qq \tp(r_c)= \frac{1}{\sqrt{g_{rr}(r_c)}}\(\frac{g_{tt}^{\prime}(r_c)}{g_{tt}(r_c)}+\frac{g_{xx}^{\prime}(r_c)}{g_{xx}(r_c)}\)+2\Cb.
\label{Eq: -0} \eeqa $\te(r_c)$ and $\tp(r_c)$ could be regarded as
the energy density and pressure of the dual fluid at rest on the
{\cutoff} surface in the thermal equilibrium state. They are relate
to each other through the equation of state of the fluid, which
comes from the Hamiltonian constraint (see Eq.~(\ref{Eq:H}) in
appendix \ref{sec:gs}).  At the zeroth order it is given by \beqa
\frac{1}{16}\(\te(r_c)+2\Cb\)\(\te(r_c)+4\tp(r_c)-6\Cb\)=\Lambda+\frac{V(\theta)}{2}+\frac{1}{2}\(\frac{1}{2}-\frac{1}{g_{rr}(r_c)}\)\theta'^2(r_c)
\eeqa The entropy density and temperature of the dual fluid is
defined as the local physical quantities at the {\cutoff} surface
$\Sigma_c$ of the black brane \cite{Bredberg:2010ky,Cai:2012vr}
\beqa \ts(r_c)=\frac{1}{4
G_N}\frac{g_{xx}(r_{h})}{g_{xx}(r_{c})},\qq
T(r_{c})=\frac{T_{H}}{\sqrt{g_{tt}(r_{c})}},\qq T_{H}\equiv
\lim_{r\to r_{h}}
\frac{g_{tt}^{\prime}(r)}{4\pi\sqrt{g_{tt}(r)g_{rr}(r)}},
\label{Eq:TH} \eeqa where a factor in the entropy density $\ts(r_c)$
has been chosen to meet with the Bekenstein-Hawking entropy formula,
while the local temperature at the {\cutoff} surface $T(r_{c})$
relates with the Hawking temperature $T_H$ through the Tolman
redshift relation.

At order $\ep^1$, only vector components appear in the dual stress
energy tensor
  \beqa \tT_{\htau i}^{(1)}=
\tT_{i\htau}^{(1)}=-\tw(r_c)\beta_{i},\qq \tw(r_c)= \te(r_c)+
\tp(r_c)=
\frac{1}{\sqrt{g_{rr}(r_c)}}\(\frac{g_{tt}^{\prime}(r_c)}{g_{tt}(r_c)}-\frac{g_{xx}^{\prime}(r_c)}{g_{xx}(r_c)}\),
\label{Eq:w0} \eeqa where $\tw(r_c)$ is the heat function per unit
volume. While at order $\epsilon^{2}$, the scalar and tensor modes
are \beqa
\tT_{\htau\htau}^{ \etwo}&=&\tw(r_c)\(\tE+\tbeta^2\)+\zetAp(r_c)\tO, \label{Eq:Tttc}\\
\tT_{ij}^{ \etwo}&=&\tw(r_c)\(\tP+\tbeta_{i}\tbeta_{j}\)-2\teta(r_c)\sigma_{ij}-2\teta_A(r_c)\sigma^A_{ij}-\tzetA(r_c)\tO\delta_{ij}, \label{Eq:Tijc}
\eeqa
The normalized  energy density and pressure perturbations are defined as
\beqa
\tE=\frac{d\te(r_c)}{dr_c~~}\frac{\dhr(\hx^\mu)}{\tw(r_c)},\qq
\tP=\frac{d\tp(r_c)}{dr_c~~}\frac{\dhr(\hx^\mu)}{\tw(r_c)},
\eeqa
where $\dhr(\hx^\mu)$ could be regarded as the thermodynamic parameter
due to the temperature perturbation.

The  transport coefficient appears at the first order gradient
expansion. In the traceless tensor modes, the \emph{normal shear
viscosity} reads \beqa \teta(r_c)\equiv
1+\sqrt\frac{g_{tt}(r_c)}{g_{rr}(r_c)}\tF^{\prime}(r_c)
=\frac{g_{xx}(r_h)}{g_{xx}(r_c)}, \label{Eq:etac}
\qq\frac{\teta(r_c)}{\ts(r_c)}=4G_N=\frac{1}{4\pi}, \eeqa
which shows that $\eta/s_0$ is {\cutoff} scale independent. While the
\emph{Hall viscosity} turns out to be
\beqa
 \tetaA(r_c)&\equiv &
\sqrt\frac{g_{tt}(r_c)}{g_{rr}(r_c)}\tFA'(r_c)
=\frac{\lambda}{2}\,\frac{g_{xx}(r_h)}{g_{xx}(r_c)}
\frac{\tphi^{\prime}(r_h)}{g_{rr}(r_h)}\frac{g_{tt}^{\prime}(r_h)}{g_{tt}(r_h)}
-\frac{\lambda}{2}\,\frac{\tw(r_c)\tphi^{\prime}(r_c)}{\sqrt{g_{rr}(r_c)}},
\label{Eq:etaA} \\
\frac{\tetaA(r_c)}{\ts(r_c)}&=&\frac{\lambda}{8
\pi}\frac{\tphi^{\prime}(r_h)}{g_{rr}(r_h)}\frac{g_{tt}^{\prime}(r_h)}{g_{tt}(r_h)}
-\frac{\lambda}{8
\pi}\frac{g_{xx}(r_c)}{g_{xx}(r_h)}\frac{\tphi^{\prime}(r_c)}{g_{rr}(r_c)}\(\frac{g_{tt}^{\prime}
(r_c)}{g_{tt}(r_c)}-\frac{g_{xx}^{\prime}(r_c)}{g_{xx}(r_c)}\).\label{Eq:etaAs}
\eeqa Thus, we see that $\eta_A/s_0$ is {\cutoff} scale dependent due
to the second term. When the cutoff surface approaches to the
horizon, namely, $r_c\rightarrow r_h$, one has
$\eta_A(r_c)/s_0(r_c)\rightarrow 0$. Even without solving the
scalar perturbation equations in (\ref{Eq:Scalarp}), we can still
give the formula of curl viscosity with the formal functions in Eq.(\ref{Eq:ds2ABCr}). In the dual
fluid stress energy tensor (\ref{Eq:Tttc}) and (\ref{Eq:Tijc}), the
scalar sector coefficients  at order $\epsilon^{2}$ are \beqa
\zetAp(r_c)&=&\frac{2}{\sqrt{g_{rr}(r_c)}}\(\frac{g_{xx}^{\prime}(r_c)}{g_{xx}(r_c)}H_{r\htau}(r_c)-H_{\hx\hx}^{\prime}(r_c)\),\label{Eq:TssHc}\\
\zetA(r_c)&=&\frac{1}{\sqrt{g_{rr}(r_c)}}\(\frac{g_{tt}^{\prime}(r_c)}{g_{tt}(r_c)}H_{r\htau}(r_c)-H_{\htau\htau}^{\prime}(r_c)\)+\frac{\zetAp(r_c)}{2}.
\eeqa Here we have used the Dirichlet boundary condition
$H_{\htau\htau}(r_c)=0$ and $H_{xx}(r_c)=0$. If the Landau frame is
chosen that $\zetAp(r_c)=0$, we can identify $\zetA(r_c)$ as the
curl viscosity of the dual fluid.

According to the Gauss-Codazi equation in pure geometry sector (see
Eq.~(\ref{Eq:GCc}) in appendix.\ref{sec:gs}) and the gravity
equations in Eq.~(\ref{Eq:EMNt}), we have \beqa
\partial^{\mu}\tT_{\mu\nu}
\equiv -2\tN^{A}R_{AB}h^{B}_{~\nu}=
 \tN^{A}\(2\lambda C_{AB}- T_{AB}^{\tphi}\)h^{B}_{~\nu}, \label{Eq:GCr}
\eeqa where $h_{AB}=g_{AB}-\tN_A \tN_B$ is the induced metric at the
{\cutoff} surface. With the gauge choice for the scalar
perturbations in Eqs.~(\ref{Eq:fullmetric})-(\ref{Eq:fullscalar}),
one can show that up to order $\ep^3$ of the source terms on the
right-handed side, the Cotton tensor $C_{AB}$ contributes nothing,
while the stress energy tensor $T_{AB}^{\tphi}$ contributes an
external force density as \beqa \mF_{i}(r_c)=- \tN^{A}
T_{AB}^{\tphi}h^{B}_{~i}=-\frac{\theta'(r_c)^2}{\sqrt{g_{rr}(r_c)}}\tpart_i\(\dhr(\hx^\mu)\),
\eeqa which comes from the scaling transformation parameter
$\delta\hr(\hx^\mu)$, and the normalized force density is
$f_i(r_c)=\mF_i(r_c)/\tw(r_c)$. Thus finally we get the
incompressible Navier-Stokes equations with Hall and curl
viscosities in Eq.~(\ref{Eq:NS}) from Einstein gravity with
Chern-Simons term, \beqa \tpart_{i}\beta^{i}=0,\quad
\tpart_{\htau}\beta_{i}+\beta_{j}\tpart^{j}\beta_{i}+\tpart_{i}\tP
-{\nu}(r_c)\,\tpart^{2}\beta_{i}-{\nu}_{A}(r_c)\epsilon^{ij}\tpart^{2}\beta_{j}-\xi_{A}(r_c)
\epsilon^{jk}\tpart_{i}\tpart_{j}\beta_{k}= f_i(r_c). \eeqa where
the kinematic viscosities are defined as \beqa
{\nu}(r_{c})={\teta(r_{c})}/{\tw(r_{c})},\quad {\nu}_{A}(r_{c})=
{\tetaA(r_{c})}/{\tw(r_{c})},\quad {\xi}_{A}(r_c)=
{\tzetA(r_c)}/{\tw(r_c)}, \label{Eq:nu} \eeqa respectively. The
Reynolds number of the dual fluid associated with the shear
viscosity and Hall viscosity can also be defined as \beqa {\mathcal
{R}} &\equiv & \frac{u L}{\nu}\propto
\frac{1}{{\nu}(r_{c})},\quad\qq {\mathcal {R}}_A\equiv  \frac{u
L}{\nu_{A}}\propto  \frac{1}{{\nu_{A}}(r_{c})}, \label{Eq:Reynolds}
\eeqa where $u$ is the characteristic velocity, $L$ is the
characteristic scale. If we further set $\dhr(\hx^\mu)\equiv0$, then
the normalized pressure perturbation $\tP$ and force density
$f_i(r_c)$ vanish, the holographic incompressible Navier-Stokes
equations will reduce into the incompressible Burger's
equations~\cite{Landau} with Hall viscosity and curl viscosity,
\beqa \tpart_{i}\beta^{i}=0,\quad
\tpart_{\htau}\beta_{i}+\beta_{j}\tpart^{j}\beta_{i}
-{\nu}(r_c)\,\tpart^{2}\beta_{i}-{\nu}_{A}(r_c)\epsilon^{ij}\tpart^{2}\beta_{j}-\xi_{A}(r_c)
\epsilon^{jk}\tpart_{i}\tpart_{j}\beta_{k}= 0. \eeqa

\section{Dual Fluid at the Infinite Boundary}
\label{sec:boundary} To compare with results in
Ref.~\cite{Saremi:2011ab}, we  take the limit with the {\cutoff}
surface $r_c\rightarrow\infty$. The ansatz of the bulk metric is
\beqa ds^{2}=-r^{2}f(r)d\tau^{2}+2H(r)d\tau dr + r^{2} dx^{i}dx_{i},
\quad i,j=1,2, \label{Eq:ds2-son} \eeqa
  $\tphi=\tphi(r)$ is a
pseudo scalar coupled to the gravitational Chern-Simons term, and
$\Lambda=-{3}/{\ell^{2}}$ is the negative cosmological constant.
Under this ansatz, the Pontryagin density $\hrR$ is identically
zero~\cite{Grumiller:2008ie}. This metric is related to our generic
metric in Eq.~(\ref{Eq:ds2-Eddington-Finkelstein}) through
 \beqa
g_{tt}(r)=r^{2}f(r),\quad g_{xx}(r)=r^{2},\quad
g_{rr}(r)=\frac{H(r)^2}{r^{2}f(r)}.
 \eeqa
 The pseudo scalar equation
$\tW^{(0)}_\theta=0$, with Eq.~(\ref{Escalar}), becomes
\beqa
\tphi^{\prime\prime}(r)+\(\frac{4}{r}+\frac{f^{\prime}(r)}{f(r)}-\frac{H^{\prime}(r)}{H(r)}\)\tphi^{\prime}(r)-\frac{H(r)^{2}}{r^{2}f(r)}\frac{dV(\tphi(r))}{d\tphi(r)}=0.
\label{Eq:EOM-theta-H-f} \eeqa
  The shear viscosity and  Hall
viscosity at the {\cutoff} surface $\Sigma_c$ defined in
Eqs.~(\ref{Eq:etac}) and (\ref{Eq:etaA}) turn out to be \beqa
\teta(r_c)=\frac{r_h}{r_c},\qq
\tetaA(r_c)=\frac{\lambda}{r_{c}^{2}}\(\frac{r_{h}^{4}f^{\prime}(r_h)\tphi^{\prime}(r_h)}{2H(r_h)^{2}}-\frac{r_{c}^{4}f^{\prime}(r_c)\tphi^{\prime}(r_c)}{2H(r_c)^{2}}\).
\eeqa
 According to Eq.~(\ref{Eq:TH}), the entropy density and
temperature of the dual fluid at {\cutoff} surface $\Sigma_c$ are
\beqa \ts(r_c)=\frac{1}{4 G_N}\frac{r_h}{r_c},\qq
T(r_{c})=\frac{T_{H}}{r_{c}\sqrt{f(r_{c})}},\qq
T_{H}=\frac{r_h^{2}f'(r_h)}{4\pi H(r_h)}. \eeqa
 Thus the shear and
Hall viscosities over entropy density are given respectively by
\beqa \frac{\teta(r_c)}{\ts(r_c)}&=&\frac{1}{4\pi}~,\quad
\frac{\tetaA(r_c)}{\ts(r_c)}=\frac{\lambda}{4\pi}\frac{1}{r_{h}^{2}}\(\frac{r_{h}^{4}f^{\prime}(r_h)\tphi^{\prime}(r_h)}{2H(r_h)^{2}}-\frac{r_{c}^{4}f^{\prime}(r_c)\tphi^{\prime}(r_c)}{2H(r_c)^{2}}\).
\eeqa where we have used normalization $16\pi G_N=1$. The shear
viscosity over entropy density of the dual fluid does not run with
the cutoff surface; this is not the case for the Hall viscosity. In
the infinity boundary limit  $r_c\to\infty$, in order to compare
with previous work, we make the assumption that
 \beqa
f(r_c)\to1-\oo({r_h^3}/{r_c^3}),\qq H(r_c)\to
1-\oo({r_h^3}/{r_c^3}), \qq\tphi(r_c)\to \oo({r_h^a}/{r_c^a}), \eeqa
when $a > 0$, we can drop the second term in $\eta_A/s$, thus
recover the result in Ref.~\cite{Saremi:2011ab} \footnote{There is a
notation difference from Ref.~\cite{Saremi:2011ab}: $\lambda_{\rm
here}= -\lambda_{\rm there}$.}. In that case, the Hall viscosity
over entropy density is entirely determined by the near horizon
region of the black brane geometry.

\section{Discussions}
\label{sec:concl}

In this paper, we have investigated the non-relativistic
hydrodynamics with Hall viscosity and curl viscosity living on a
finite {\cutoff} surface, dual to $(3+1)$-dimensional Einstein
gravity with a pseudo scalar coupled to a topological gravitational
Chern-Simons term. The topological Pontryagin density $\tilde{R}R$
is totally depicted by the Weyl tensor, which describes the
traceless part of the Riemann tensor. The Ricci tensor is the
Riemann tensor's trace, and together with the Weyl tensor, provides
a complete description of the curvature of the space-time. In
(3+1)-dimensional space-time, the Weyl tensor and its dual can be
written as
 \beqa
C_{ABCD} &=& R_{ABCD}-(g_{A[C}R_{D]B}-g_{A[D}R_{C]B})+\frac{1}{3}g_{A[C}g_{D]B}R,\\
\tilde{C}^{ABCD}&=& \frac{1}{2}\epsilon^{CDEF}C^{AB}_{\ \
\,\,\,\,EF},\qq \tilde{C}C\equiv\tilde{C}^{ABCD}C_{BACD}, \eeqa and
it can be shown that the Pontryagin density $\tilde{R}R=\tilde{C}C$
(for example see~\cite{Grumiller:2007rv}). One can further define
the gravito-electric and gravito-magnetic field as \beqa E_{AC} + i
B_{AC} \equiv (C_{ABCD}+i \tilde{C}_{ABCD})u^{B}u^{D}=(C_{ABCD}+i
\tilde{C}_{ABCD})P^{BD}, \eeqa where $u^{B}$ is a normalized
time-like 4-velocity, and its projection $P^{BD}=g^{BD}+u^B u^D$.
The last identity is true due to the traceless properties of the
Weyl tensor. Thus, the Lagrangian density in
Eq.~(\ref{Eq:LCS-theta}) in the bulk can be expressed as
 \beqa {\mathcal L}_{CS}\equiv\frac{\lambda}{4}\,\theta
\tilde{R}R=\frac{\lambda}{4}\theta\,\tilde{C}C=-4{\lambda}\,\theta\,E_{AC}B^{AC},
\eeqa which is analogous to that in $U(1)$ gauge theory, ${\mathcal
L}_{CS}^{(\kappa)}=\kappa
\theta\tilde{F}^{\mu\nu}F_{\mu\nu}=-4\kappa \theta (\vec{E}\cdot
\vec{B})$. With suitable coefficient and a constant $\theta$-vacuum,
${\mathcal L}_{CS}^{(\kappa)}$ can be used to describe the
electromagnetic response of (3+1)-dimensional topological insulators
(TIs), while ${\mathcal L}_{CS}$ can be used to describe the
gravitational response of (3+1)-dimensional topological
superconductors (TSCs), a half quantized thermal Hall effect may
appear on the surface~\cite{Ryu:2010,Wang:2011}. Alternatively, if
we consider the stress response of (3+1)-dimensional TIs to an
external torsion field, it was shown in Ref.~\cite{Hughes:2011hv}
that the time-reversal invariant TIs will exhibit a quantum Hall
viscosity on their surfaces. Thus it would be interesting to see
whether the bulk gravity in this paper could describe a deformed TIs
or TSCs, with the surface quantum Hall effect as the dual
description. In our general perturbative metric (\ref{Eq:ds2-nc}),
the Lagrangian of the anomaly term firstly appears at order $\ep^2$
as \beqa\label{Eq:Lcs} {\mathcal
L}_{CS}=\frac{\lambda}{2}\,\theta(r)\frac{g_{xx}(r_c)}{g_{rr}^{{3}/{2}}(r)}
\frac{g_{xx}^{\prime}(r)}{g^2_{xx}(r)}\sqrt\frac{g_{tt}(r)}
{g_{tt}(r_c)}\(\frac{g_{tt}^{\prime}(r)}{g_{tt}(r)}-\frac{g_{xx}^{\prime}(r)}{g_{xx}(r)}\)^{2}\tO,
\eeqa which has been simplified through the background equations in
(\ref{Eq:Bg}). To get a non-vanishing Hall viscosity $\etaA$ in
Eq.~(\ref{Eq:etaA}), the pseudo scalar field $\tphi(r)$ is required
to be coordinate $r$-dependent. Otherwise, when $\tphi(r)$ vanishes,
parity is not broken in the bulk and in the dual boundary theory,
and when $\tphi(r)$ is a non-zero constant, the gravitational
Chern-Simons term is just a surface term in the action which
violates the parity in the bulk but contributes nothing to the
equations of motion in the bulk, hence leads to a vanishing Hall
viscosity. In Ref.~\cite{Chen:2011fs}, by including a gauge field in
the bulk gravity such that the solution is free from the violation
of the positive energy theorem~\cite{Hertog:2006rr}, the pseudo
scalar hair would break parity spontaneously.

Inheriting the spirit of holographic Wilson's renormalization group
(RG) approach in Ref.~\cite{Bredberg:2010ky}, the re-scaled
{\cutoff} size could be regarded as the RG running scale. The
non-vanishing trace of the symmetric stress energy tensor leads to
the trace anomaly in quantum field theory, where the corresponding
$\beta$-function can be obtained from the coefficients in front of
the trace. In our dual gravity model, Setting $\lambda=0$ to turn
off the gravitational Chern-Simons term in the bulk, we can go back
to the d-dimensional fluid corresponding to (d+1)-dimensional bulk
gravity with a scalar field $\theta(r)$, the relation between the
trace $\tT^{\mu}_{~\mu}$ in Eq.~(\ref{Eq:Trace}) of dual Brown-York
stress energy tensor at zeroth order and $\beta_{( e)}$ is, \beqa
\tT^{\mu}_{~\mu}=-\beta_{(e)}\frac{d}{r_c\[g^{d/2}_{xx}(r_c)\]'}
+\frac{2g_{xx}(r_c)}{g'_{xx}(r_c)}\frac{\theta'(r_c)^2}{\sqrt{g_{rr}(r_c)}}\,,
\qq\beta_{( e)}= r_c\frac{\partial}{\partial  r_c}\[{
e(r_c)}g^{d/2}_{xx}(r_c)\],
 \eeqa
  where $ e(r_c)$ is given in
Eq.~(\ref{Eq:d-e}), $r_c$ is analogous to the {\cutoff} energy
scale. Considering there exists a conformal factor difference in the
definition of the dual stress tensor in Ref.~\cite{Brattan:2011my}
and transforming to that frame, we can conclude that the trace of
the fluid stress tensor is generated by the energy density as well
as an extra term contributed by  the scalar field. This relation
could also be generalized at higher order of hydrodynamic expansion,
and in order to consider the effects of parity violating
interactions, one can turn on the gravitational Chern-Simons term in
the bulk. As the ratio of the Hall viscosity over entropy density in
Eq.~(\ref{Eq:etaAs}) is found to be cutoff scale dependent, we can
also define the following $\beta$-function
\beqa
 \beta_{(\eta_A/s)}
=r_c\frac{ \partial }{\partial r_c }\[\frac{\eta_A(r_c)}{\ts(r_c)}\]
=-\frac{\lambda\, r_c}{8 \pi}\frac{ \partial }{\partial r_c
}\[\frac{g_{xx}(r_c)}{g_{xx}(r_h)}\frac{\tw(r_c)\tphi^{\prime}(r_c)}{\sqrt{g_{rr}(r_c)}}\],
\eeqa
 to represent its non-trivial evolution, where $\tw(r_c)$ is
the heat function per unit volume in Eq.~(\ref{Eq:w0}). When the
cutoff surface approaches to the horizon of the background
spacetime, $\eta_A(r_h)/s_0(r_h)=0$, but $\beta_{(\eta_A/s)}$ could
be non-zero.

Note that in our model, the ratio of Hall viscosity to entropy
density depends on the cutoff scale, while the ratio of shear
viscosity to entropy density does not. This means that the shear
viscosity has a same dependence of the cutoff as the entropy
density, but it is not for the Hall viscosity. Thus it is of some
interest to ask whether one can construct a quantity involving the
Hall viscosity, which is cutoff independent. While having considered the form (\ref{Eq:etaA}), at the moment we have no
idea on the Hall viscosity. Here we just mention that a similar phenomena appears in charged fluid
with anomalous current~\cite{Son:2009tf}. It was recently shown that the anomaly vortical coefficient $\xi$
depends on the cutoff scale, while the ratio of $\xi$ to a function of thermodynamic quantities is cutoff
independent~\cite{Bai:2012ci}. A similar denominator for the curl viscosity is expected to be found,
but the approach does not applies to the Hall viscosity since it is non-dissipative.
Certainly it is required to further understand the dependence of the cutoff for the ratio of Hall
viscosity over entropy density in the parity broken fluid~\cite{Jensen:2011xb} or weak-isospin incompressible quantum liquid~\cite{Perelstein:2010yd}.

\vskip1cm

\section*{Acknowledgements}

We thank Song He, Ya-Peng Hu, Li Li, Jing Ren, Jia-Rui Sun,
Xu-Feng Wang, Zhong-Zhi Xianyu, Jia-Jun Xu and Yang Zhang for 
various valuable discussions. This work is supported
in part by the National Natural Science Foundation of China (No.
10821504, No. 10975168, No. 11035008, No. 11075194 and No.
11135003), and in part by the Ministry of Science and Technology of
China under Grant No. 2010CB833004. Particularly, Y.~-H.~Qi would
like to thank Yu-Ping Kuang and S.-S.-Henry Tye for advice and
support, he also would like to thank \'Eanna Flanagan for instructing
him general relativity. Y.~-L.~Zhang would like to thank the
``Spring School on Superstring Theory and Related Topics'', held at
the Abdus Salam International Centre for Theoretical Physics (ICTP:
smr2331), for the hospitality and financial support, and H.~Liu and
S.~Minwalla for kind and valuable discussions during the school.

\appendix

\section{Non-relativistic Hydrodynamic from Gravity}

\label{sec:gs} This appendix briefly introduces the non-relativistic
expansion procedure associated with a {\cutoff}
surface~\cite{Cai:2012vr}.
For further details, see Refs.~\cite{Bhattacharyya:2008kq,Bredberg:2011jq,Bredberg:2010ky,Brattan:2011my,Iqbal:2008by,Nickel:2010pr,
Compere:2011dx,Cai:2011xv,Lysov:2011xx,Niu:2011gu}.

\subsection{Bulk Geometry and
a Finite {\Cutoff} Surface} \label{subsec:IG}
In order to study the fluid in $d$-dimensional flat space-time, let
us consider a $(d+1)$-dimensional background geometry with a generic
metric \beqa
ds^{2}_{d+1}&=&-g_{tt}(r)dt^{2}+g_{xx}(r)\delta_{ij}dx^{i}dx^{j}+g_{rr}(r)dr^{2},\quad
i,j=1,\ldots d-1,\label{Eq:ds2-generic1} \eeqa where the metric
components depend only on radial coordinate $r$.
 In addition, we assume the fully spatial rotational $SO(d-1)$ symmetry in $x^{i}$ directions, and require that
\emph{the geometry of the space-time manifold ${\mathcal M}$ has a
well-defined future horizon located at $r=r_{h}$,} where $g_{tt}(r)$
has the first-order zero $g_{tt}(r_h)=0$, and $g_{rr}(r)$  has the
first-order pole $g^{-1}_{rr}(r_h)=0$ ~\cite{Iqbal:2008by}. By using
the Eddington-Finkelstein coordinate $\tau$ defined through $d\tau =
dt +\sqrt{{g_{rr}(r)}/{g_{tt}(r)}}\,dr$, we can rewrite the bulk
metric as \beqa
ds^{2}_{d+1}=-g_{tt}(r)d\tau^{2}+g_{xx}(r)dx_{i}dx^{i}+2\sqrt{g_{tt}(r)g_{rr}(r)}d\tau
dr. \label{Eq:ds2-Eddington-Finkelstein} \eeqa

Let us define an arbitrary hyper-surface at a constant radial
coordinate $r$ as $\Sigma_r$, and introduce a specific finite
{\cutoff} hyper-surface  $\Sigma_c$ at $r=r_{c}$ outside the horizon
($r_{c}>r_{h}$), where the associated intrinsic coordinates
$\hat{x}^{\mu}\sim (\hat{\tau}, \hat{x}^{i})$ on $\Sigma_c$ are
\beqa \hx^{0}=\hat{\tau}=\tau\sqrt{g_{tt}(r_{c})}, \quad
\hx^{i}={x}^{i}\sqrt{g_{xx}(r_{c})},\quad i= 1,2,\ldots, d-1. \eeqa
The bulk metric in Eq.~(\ref{Eq:ds2-Eddington-Finkelstein}) in the
intrinsic coordinates becomes \beqa
d\hs^{2}_{d+1}=-\frac{g_{tt}(r)}{g_{tt}(r_{c})}d\hat{\tau}^{2}+\frac{g_{xx}(r)}{g_{xx}(r_{c})}d\hat{x}_{i}d\hat{x}^{i}+2\sqrt{\frac{g_{tt}(r)}{g_{tt}(r_{c})}}\sqrt{g_{rr}(r)}d\hat{\tau}
dr. \label{BulkMetric} \eeqa We adopt the Arnowitt-Deser-Misner(ADM)
decomposition along $r$ direction~\cite{ADW} \beqa
ds^{2}_{d+1}=\alpha^{2}(r)dr^{2}+{\gamma}_{\mu\nu}(r)(d\hat{x}^{\mu}-\beta^{\mu}(r)dr)(d\hat{x}^{\nu}-\beta^{\nu}(r)dr),\label{ADM}
\eeqa with the \emph{``lapse function''}
$\alpha(r)=\sqrt{g_{rr}(r)}$ and the
 \emph{``shift vector''} $\beta^{\mu}(r)= \sqrt{{g_{tt}(r_{c})}/{g_{tt}(r)}} \sqrt{g_{rr}(r)}\,\delta^{\mu}_{\tau}$ respectively.
The generic \emph{induced metric} $\gamma_{\mu\nu}(r)$ is an
analytically extension from $\Sigma_{c}$ to $\Sigma_r$, where \beqa
\gamma_{\mu\nu}(r)d\hat{x}^{\mu}d\hat{x}^{\nu}\equiv
-\frac{g_{tt}(r)}{g_{tt}(r_{c})}d\hat{\tau}^{2}+\frac{g_{xx}(r)}{g_{xx}(r_{c})}d\hat{x}_{i}d\hat{x}^{i}.
\label{InducedMetric} \eeqa It is worthy to notice that
${\gamma}_{\mu\nu}(r_{c}) =\eta_{\mu\nu}\equiv(-1,\delta_{ij})
\label{{\Cutoff}Metric}$, which implies that the induced metric
$\gamma_{\mu\nu}(r)$ on $\Sigma_r$ reduces to a flat metric on
$\Sigma_c$.
\subsection{Diffeomorphisms Associated
 with the {\Cutoff} Surface}

With a flat induced metric given at the {\cutoff} surface
$\Sigma_{c}$, we can take into consider two finite diffeomorphisms,
which preserve the induced metric invariant on the cutoff surface.
The first one is a linear scale transformation along the radial
coordinate $r$ and the rescaling of the intrinsic coordinates
$\hat{x}^{\mu}=(\hat\tau,\hat{x}^{i})$, \beqa r\to \hr\equiv k(r),
\quad \hat{\tau}\to
\hat{\tau}\sqrt{\frac{g_{tt}(r_{c})}{g_{tt}(\hr_{c})}}, \quad
\hat{x}^{i}\to
\hat{x}^{i}\sqrt{\frac{g_{xx}(r_{c})}{g_{xx}(\hr_{c})}},
\label{Scale-Trans} \eeqa where $k(r)$ is a linear function of $r$,
and a concrete form of $k(r)$ will be chosen according to the specific global geometry of the bulk metric in Eq.~(\ref{BulkMetric}). In addition, 
the notation $\hr_{c}\equiv k(r_{c})$ is introduced. Under this
diffeomorphism, the bulk metric in Eq.~(\ref{BulkMetric}) is
rescaled as \beqa
d\hs^{2}_{d+1}=-\frac{g_{tt}(\hr)}{g_{tt}(\hr_{c})}d\htau^{2}
+\frac{g_{xx}(\hr)}{g_{xx}(\hr_{c})}d\hx_{i}d\hx^{i}+2\sqrt{\frac{g_{tt}(\hr)}{g_{tt}(\hr_{c})}}\sqrt{g_{rr}(\hr)}d\htau
d\hr. \label{BulkMetric_Scaled}\eeqa
The second diffeomorphism is a generic Lorentz transformation \beqa
\hat{\tau}\to \gamma(\hat{\tau}-\beta_{i}\hat{x}^{i}),\quad
\hat{x}^{i}\to
\hat{x}^{i}-\gamma\beta^{i}\hat{\tau}+(\gamma-1)\frac{\beta^{i}\beta_{j}}{\beta^{2}}\hat{x}^{j},
\eeqa with the boost parameter $\beta_{j}=\delta_{ij}\beta^{i}$ and
Lorentz factor $\gamma=1/\sqrt{1-\beta^{2}}$. We can define the
\emph{$d$-dimensional velocity} and \emph{$d$-dimensional
polarization vector} as \beqa u_{\mu} &=&\gamma(-1,\beta_{i}), \quad
n_{\mu}^{j}=(-\gamma\beta^{j},~\delta^{j}_{i}+(\gamma-1)\frac{\beta^{j}\beta_{i}}{\beta^{2}}),\label{Eq:um}
\eeqa with $u^{\mu}u_{\mu}=-1$, $n_{j}^{\mu}n_{\mu}^{j}=d-1$ and
$u^{\mu}n_{\mu}^{i}=0$. Then the Lorentz transformation on the
{\cutoff} hypersurface becomes $\hat{\tau}\to
-\hu_{\mu}\hat{x}^{\mu}$, and $\hat{x}^{i}\to
\hn_{\mu}^{i}\hat{x}^{\mu}$.
The re-scaled bulk metric in Eq.~(\ref{BulkMetric_Scaled}) becomes
\beqa  d\hs
^{2}_{d+1}=-\frac{g_{tt}(\hr)}{g_{tt}(\hr_{c})}\hu_{\mu}\hu_{\nu}d\hat{x}^{\mu}d\hat{x}^{\nu}+\frac{g_{xx}(\hr)}{g_{xx}(\hr_{c})}
{P}_{\mu\nu}d\hat{x}^{\mu}d\hat{x}^{\nu}-2
\sqrt{\frac{g_{tt}(\hr)}{g_{tt}(\hr_{c})}}\sqrt{g_{rr}(\hr)}\hu_{\mu}d\hat{x}^{\mu}d\hr,
\label{Eq:BM}\eeqa
where the \emph{projection tensor}
\beqa
{P}_{\mu\nu}\equiv\eta_{\mu\nu}+\hu_{\mu}\hu_{\nu}=\delta_{ij}\hn_{\mu}^{i}\hn_{\nu}^{j}
=\left(
                       \begin{array}{cc}
                       \gamma^{2}\beta^{2} & -\gamma^{2}\beta_{j} \\
                       -\gamma^{2}\beta^{i} & \delta^{i}_{~j}+\gamma^{2}\beta^{i}\beta_{j} \\
                       \end{array}
                     \right), \label{Eq:Pmn}
\eeqa
satisfying ${P}^{\mu}_{\mu}=d-1$.
Under the two diffeomorphism transformations, the metric in the 
ADM-like decomposition with the ``lapse'' function $\halpha(r)$ and the ``shift'' vector $\hbeta(r)$ becomes
\beqa
d\hs^{2}_{d+1}&=& \halpha^{2}(r)d\hr^{2}+\hgamma_{\mu\nu}(r)\(d\hat{x}^{\mu}-
\hbeta^{\mu}(r)d\hr\)\(d\hat{x}^{\nu}-\hbeta^{\nu}(r)d\hr\), \label{ADM_Diffeomorphisms} \\
\halpha(r)&=&\sqrt{g_{rr}(\hr)},\quad
\hbeta^{\mu}(r)=\sqrt{\frac{g_{tt}(\hr_{c})}{g_{tt}(\hr)}}
\sqrt{g_{rr}(\hr)}~\hu^{\mu}, \eeqa and the induced metric in
Eq.~(\ref{InducedMetric}) is transformed into $\hgamma_{\mu\nu}(r)$
as \beqa {\hgamma}_{\mu\nu}(r)\equiv
-\frac{g_{tt}(\hr)}{g_{tt}(\hr_{c})}\hu_{\mu}\hu_{\nu}+\frac{g_{xx}(\hr)}{g_{xx}(\hr_{c})}{P}_{\mu\nu}
\label{InducedMetric_Scaled_Lorentz}. \eeqa At the {\cutoff}
$r=r_{c}$ surface, which is equivalent to  $\hr=\hr_{c}$, the
induced metric $\hgamma_{\mu\nu} (r_c)$   keeps flat.

\subsection{Dual Hydrodynamic in Non-Relativistic Limit}

\label{subsec:nonrel}

 We perturb the diffeomorphism metric (\ref{Eq:BM}) in
the non-relativistic limit. As $r$ is the radial coordinate along
the extra dimension orthogonal to the {\cutoff} hyper-surface, we
can define a scaling transformation parameter $\delta \hr\equiv
\hr-r$. Assuming the scalar parameter $\delta \hr$ and the velocity
parameters $\beta^{i}$ in the Lorentz boost are all functions of the
intrinsic coordinates $\hx^{\mu}=(\htau,\hx^{i})$, in the
non-relativistic hydrodynamic limit we have the following
scalings\,(\,here the subscript $i$ present $\hat{x}^{i}$) \beqa
\partial_{r}\sim \epsilon^{0}, \quad \tpart_{i}\sim\tbeta_{i}(\hx^\mu)\sim \epsilon^{1},\quad
\tpart_{\htau}\sim \delta \hr(\hx^\mu)\sim\epsilon^{2},
\label{Eq:HDE} \eeqa where $\ep\ll 1$ is a small parameter.
Therefore, the metric function can be expanded, i.e.,
$g_{tt}(\hr)=g_{tt}(r)+g_{tt}^{\prime}(r)\delta \hr(r)+\ldots$, then
the diffeomorphism metric can be expanded
order by order in the expansion parameter, 
up to order $\ep^2$, which gives the result in
Eq.~(\ref{Eq:ds2-nc}).
In Einstein gravity, we assume that the generic metric in
(\ref{Eq:ds2-generic1}) solves Einstein's equations. Then the
diffeomorphism metric is also the solution. After demanding the
transformation parameters $\tbeta_{i}$ and $\delta\hr$ be
coordinates $\hx^\mu$ dependent, the metric in (\ref{Eq:BM}) is no
longer the solution of Einstein's equations. We can solve the
Einstein's equations order by order in the non-relativistic
hydrodynamic expansion, via adding new correction terms to the
metric.  Similar procedure is also applicable in the case when
matter fields are present.

The \emph{intrinsic curvature} of the bulk geometry is measured by
Riemann tensor, while the \emph{extrinsic curvature} depends on how
the hyper-surface is embedded into the bulk geometry. Thus, when we
perturb the background metric, we can choose the gauge to keep the
induced metric $\hgamma_{\mu\nu}(r_c)$ flat. The extrinsic curvature
at the {\cutoff} surface  is $\hK_{\mu\nu}=\frac{1}{2}{\mathcal
L}_{\tN}\hgamma^{(c)}_{\mu\nu}(r)|_{r=r_c}$, where ${\mathcal
L}_{\tN}$ is the Lie derivative along the outpointing normal vector
$\tN$ of the {\cutoff} surface, $\hgamma^{(c)}_{\mu\nu}(r)$ is an
analytic extension of the induced metric $\hgamma_{\mu\nu}(r_c)$ on
$\Sigma_{c}$. And they are all associated with the perturbed metric
with correction terms.
In $(d+1)$-dimensional Einstein gravity, if we use the unit $16\pi
G_N ^{(d+1)}=1$, the Brown-York tensor on the {\cutoff} surface
$\Sigma_c$ is given by~\cite{Brown:1992br}
\beqa
\tT^{BY}_{\mu\nu}=2(\hK\hgamma_{\mu\nu}(r_c)-\hK_{\mu\nu}+{C}\hgamma_{\mu\nu}(r_c)).\label{Eq:BYr}
\eeqa It can be identified with the stress tensor of the dual fluid
 Ref.~\cite{Bredberg:2010ky}.
 In the non-relativistic
hydrodynamic expansion, we can also fix the physical stress energy
tensor order by order. The zeroth order stress tensor can be written
as $\tT_{\mu\nu}^{BY}= e(r_c)u^\mu u^\nu + p(r_c)P_{\mu\nu}$,
where \beqa
\qq~~~ e(r_c)&=&-\frac{d-1}{\sqrt{g_{rr}(r_c)}}\frac{g_{xx}^{\prime}(r_c)}{g_{xx}(r_c)}-2\Cb,\label{Eq:d-e}\\
\qq
p(r_c)&=&\frac{1}{\sqrt{g_{rr}(r_c)}}\(\frac{g_{tt}^{\prime}(r_c)}{g_{tt}(r_c)}+(d-2)\frac{g_{xx}^{\prime}(r_c)}{g_{xx}(r_c)}\)+2\Cb,
\label{Eq:d-p} \eeqa are the energy density and pressure of the dual
fluid, respectively. And the trace of the dual stress energy tensor
at the zeroth order is \beqa \tT^{\mu}_{~\mu}=- e(r_c)+(d-1)p(r_c). \label{Eq:Trace}\eeqa
 In addition, the dual fluid is described by the constraint
equations~\cite{Lysov:2011xx,Wald:1984rg} \beqa
(\hK^2-\hK_{AB}\hK^{AB})&\equiv & 2\, G_{MN} \tN^{M}\tN^{N}|_{\Sigma_c},\label{Eq:H}\\
2D^{A}(\hK h_{AB}-\hK_{AB})&\equiv &-2 G_{MN}
\tN^{M}h^{N}_{~B}|_{\Sigma_c}. \label{Eq:GCc} \eeqa where $G_{MN}$
is the Einstein tensor, $h_{AB}=g_{AB}-\tN_A\tN_B$ and $D$ denotes
the derivative operator associated with the induced metric $h_{AB}$.
The first one is the Hamiltonian constraint which gives the equation
of state of fluid relating the pressure and energy density. The
second one is the momentum constraint which gives the evolution
equations of the fluid, and it reduces to the incompressible
Navier-Stokes equations in the non-relativistic limit. Correction
terms will appear if matters or higher order curvature terms are
added to Einstein gravity. For example, the matters will provide
external sources to the equations of motion, and the Gauss-Bonnet
term will lead to an additional term for the shear
viscosity~\cite{Cai:2012vr}.

\bibliography{000}

\end{document}